\documentclass[useAMS,usenatbib]{mn2e}

\usepackage{graphicx,enumerate,amssymb,url,perpage,aas_macros,float}
\def\lesssim{\la}
\def\gtrsim{\ga}

\title[Searching for Eclipsing Disks]{A search for eclipsing binaries that host discs}

\author[Z.~Meng et al.]{Zeyang Meng$^{1,2}$\thanks{E-mail: mengzy1989@gmail.com}, Alice C. Quillen$^2$, Cameron P. M. Bell$^{2,3}$, 
\newauthor
Eric E. Mamajek$^2$, Erin L. Scott$^2$, Ji-Lin Zhou$^1$
\\
$^1$ Department of Astronomy \& Key Laboratory of Modern Astronomy and Astrophysics in Ministry of Education,\\Nanjing University, Nanjing, 210093, China \\
$^2$ Department of Physics and Astronomy, University of Rochester, Rochester, NY 14627, USA\\
$^3$ School of Physics, University of Exeter, Exeter EX4 4QL, UK\\
}

\begin{document}
\maketitle

\begin{abstract}
We search for systems hosting eclipsing discs using a complete sample of eclipsing binaries (EBs); 
those previously identified in the third phase of the Optical Gravitational Lensing Experiment (OGLE-III).
Within a subsample of 2,823 high-cadence, high-photometric precision and large eclipsing depth detached
EBs previously identified in the Large Magellanic Cloud (LMC), 
we find that the skewness and kurtosis of the light curves magnitude distribution within the primary eclipse 
can distinguish EBs with a complex shaped eclipse from those without. 
Two systems with previously identified eclipsing discs (OGLE-LMC-ECL-11893 and OGLE-LMC-ECL-17782) are 
identified with near zero skewness ($|S|<0.5$)  and positive kurtosis. 
No additional eclipsing disc systems were found in  the OGLE-III LMC, Small Magellanic Cloud (SMC) 
or Galactic Disc (GD) EB light curves.

We estimate that the fraction of detached early-type LMC EBs 
(which have a primary with an $I$-band magnitude brighter than $\simeq 19\,\rm{mag}$) that exhibit atypical
eclipses and so could host a disc is approximately 1/1000. 
As circumstellar disc lifetimes are short,  we expected to primarily find eclipsing discs
around young stars. In addition, as there is more room for a disc in a widely separated binary
and because a disk close to a luminous star would be above the dust sublimation temperature, 
we expected to primarily find eclipsing discs
in long period binaries.   However, OGLE-LMC-ECL-17782
is a 13.3 day period B star system with a transient and hot ($\sim 6000\,\rm{K}$, $\sim 0.1\,\rm{AU}$ radius) disc
and Scott et al. (in prep.) estimate an  age of $150\,\rm{Myr}$ for OGLE-LMC-ECL-11893. 
Both discs are unexpected in the EB sample and impel explanation.



\end{abstract}

\section{Introduction}

One  star in a young stellar binary system could host a circumstellar disc and  the disc
could periodically occult the other star as seen from a distance.
Because discs can be large,
the probability that a randomly oriented system exhibits eclipses may be fairly high 
 \citep{mamajek12}.  
Discs in binary systems could be seen in eclipse during the epoch of planet formation (in the case of a companion
circumstellar disc; \citealt{galan12}) or during the epoch of satellite formation
(in the case of a circumplanetary disc; \citealt{mamajek12}).
Disc transits can provide unique information about disc opacity and structure on scales that
are difficult to observe in any other way (e.g. on occultations of Saturn's rings; \citealt{hedman07}).

Two well-known long period and bright eclipsing systems have been interpreted in terms of occulting dark discs,
$\epsilon$ Aurigae \citep{guinan02,kloppenborg10,chadima11} and
EE Cephei  \citep{mikolajewski99,graczyk03,mikolajewski05,galan12}.
Recently \citet{mamajek12} reported a single 
long, deep, and complex eclipse event on an approximately solar mass pre-main-sequence star.

Among the eclipsing binaries (EBs) in the Large Magellanic Cloud (LMC), using the OGLE-III survey \citep{udalski08},
\citet{graczyk11} discovered an object with $13.3\,\rm{day}$ period with a semitransparent and variable disc-like structure;
 OGLE-LMC-ECL-17782.   
A second object with an eclipsing disc but with a $468\,\rm{day}$ period
 OGLE-LMC-ECL-11893 has also been identified in the same EB sample
\citep{dong14}.
OGLE-LMC-ECL-17782   was previously identified as a detached 
eclipsing binary by \citet{derekas07}.
In this paper we explore a way to automatically identify these eclipsing disc systems. 
We also search the OGLE-III LMC, Small Magellanic Cloud (SMC) and Galactic Disc (GD) EB samples \citep{graczyk11,Pawlak13,Pietrukowicz13}
for additional eclipsing disc candidates.

Recent infrared and millimeter surveys of nearby stars and star-forming regions find that
age, stellar mass and multiplicity  
affect the likelihood that a stellar system exhibits a detectable disc 
\citep{haisch01,bouwman06,hernandez07,harris12,derosa13}.
As illustrated in OGLE-III EB studies \citep{graczyk11,Pawlak13,Pietrukowicz13}, EBs detected from large
photometric surveys are a well-defined, nearly complete sample.
Following our search for EBs that host discs in light curves, we compare the fraction 
of discs found by their occultations seen in light curves to the fraction that would be expected 
as inferred from statistics of disc detection in recent Galactic infrared and millimetre surveys of lower mass stars.  
Our search for eclipsing discs in the LMC EB sample is a complimentary or
unique way to search for circumstellar discs.

In Section~\ref{searching_for_discs} we characterise
 the photometric properties of each eclipsing binary light curve 
 and we reduce the EB sample to a high-cadence, low-noise subset.
In this subset we show that the systems with disc candidates stand out in skewness vs. kurtosis space
(computed from moments of the magnitude distributions within eclipse).  
We apply our method to search the OGLE-III LMC, SMC and GD EB samples for additional eclipsing discs. 
In Section~\ref{properties_primary_stars} we  discuss the nature of the two disc-hosting systems. Section~\ref{speculation_fraction_discs} discusses the fraction of objects that host eclipsing discs based on infrared and millimetre Galactic surveys of young etars.
Finally, Section~\ref{conclusions} discusses and summarises the main findings of the paper.


\section{Searching for eclipsing discs in EB light curves}
\label{searching_for_discs}

\citet{graczyk10} illustrate
that the moments of the light curve magnitude distribution or
light curve statistical moments can be used to classify variable
stars.  This technique is particularly effective at identifying EBs and differentiating them from other
types of variable stars.
Here we apply a similar technique but only to the region of the light curve within eclipse.
Our goal is to use light curves to automatically identify eclipsing disc candidate systems previously  identified as EB systems.

In eclipsing binaries, a star passing in front or behind another star,
gives a triangular, square or Gaussian  shaped feature in the light curve.
Here we are searching for eclipse shapes (features in the light curve) that differ from the expected shapes.
We are searching for asymmetric transits (such as seen in the light curve of EE Cep; \citealt{galan12} or
J1407 \citealt{mamajek12}), a W-shaped eclipse (such as in $\epsilon$ Aurigae; \citealt{guinan02})
or a stellar transit bracketed by a wider, lower depth platform (such as OGLE-LMC-ECL-17782; \citealt{graczyk11}).
If we find such an unusual shape in a periodic light curve we label it as an eclipsing disk candidate.
Associated rotation (implying that they are in fact disks) has not yet been seen in these 
objects.
 
 \subsection{Method}

A brief overview of our procedure to identify possible EB disc systems is given below, 
and then we provide an example of this procedure using the OGLE-III LMC, SMC and GD EB samples.  

\begin{enumerate}[(i)]
\item
Using a phase-folded light curve of an object previously identified as a detached eclipsing binary in the OGLE-III survey,
we measure the mean magnitude and dispersion of photometric points outside of eclipse. 
\item
We discard noisy systems with too few light curve data points and with large dispersions (possibly due to variability).
\item
We identify a beginning and an end phase for the primary and secondary eclipses, thus defining eclipse windows.
\item
We measure the skewness and kurtosis of the distribution of magnitudes using photometric points within a given eclipse window.
\end{enumerate}
\noindent If noisy systems and systems with too few light curve data points are not discarded it is difficult
to identify the eclipse window. This leads to imprecise kurtosis and skewness derivations in the eclipse magnitude distribution, resulting in too many potential eclipsing disc contaminants.

Below we describe how to characterise the mean magnitude and standard deviation  of these EB systems (outside of eclipse) as well as how to identify the ingress and egress of both primary and secondary eclipses. 
 
\begin{enumerate}[(a)]
\item
We folded the OGLE-III LMC, SMC and GD light curves using the periods previously computed by \citet{graczyk11}, \citet{Pawlak13} and \citet{Pietrukowicz13} respectively.
\item 
We compute the median value of the entire light curve (points inside and outside of eclipse), $\mu_L$. Because it is a median, this should approximately be 
an average magnitude outside of eclipse.
\item 
We smooth the light curve using a box with a width of 5 data points and mark the  faintest point of the smoothed
light curve as the centre of the primary eclipse. The phase of the primary eclipse is denoted $\theta_p$. 
The phase of the secondary eclipse centre, $\theta_s$, is identified with the phase where the magnitude difference 
$|m(\theta_s) - m(\theta_m)|$ is largest. Here $\theta_m$ is the phase midway between the primary and secondary eclipses and $m(\theta)$ is the
magnitude as a function of phase.
\item 
Within a window centred at $\theta_m$ and with absolute phase width of 0.1, we estimate the standard deviation (or equivalently dispersion) $\sigma_L$ of the light curve outside of eclipse.
\item 
We define a primary in-eclipse window where the magnitude is $1\sigma$ fainter than the mean (i.e. $m > \mu_L + \sigma_L$) and contains $\theta_p$. We also define a secondary in-eclipse window in the same way containing $\theta_s$.
\item 
After masking the two eclipse windows we recalculate the median magnitude and dispersion, thereby providing a more realistic determination of these values outside of eclipse. 
With the updated values for $\mu_L$ and $\sigma_L$ we recompute the eclipse windows using the same
criterion as given in (e).
We note that the recalculated $\mu_L$ and $\sigma_L$ are 
used to calculate the kurtosis and skewness parameters within the eclipse windows.
\end{enumerate}

\begin{figure}
\includegraphics[width=\columnwidth, trim= 0.5in 0 0 0 ]{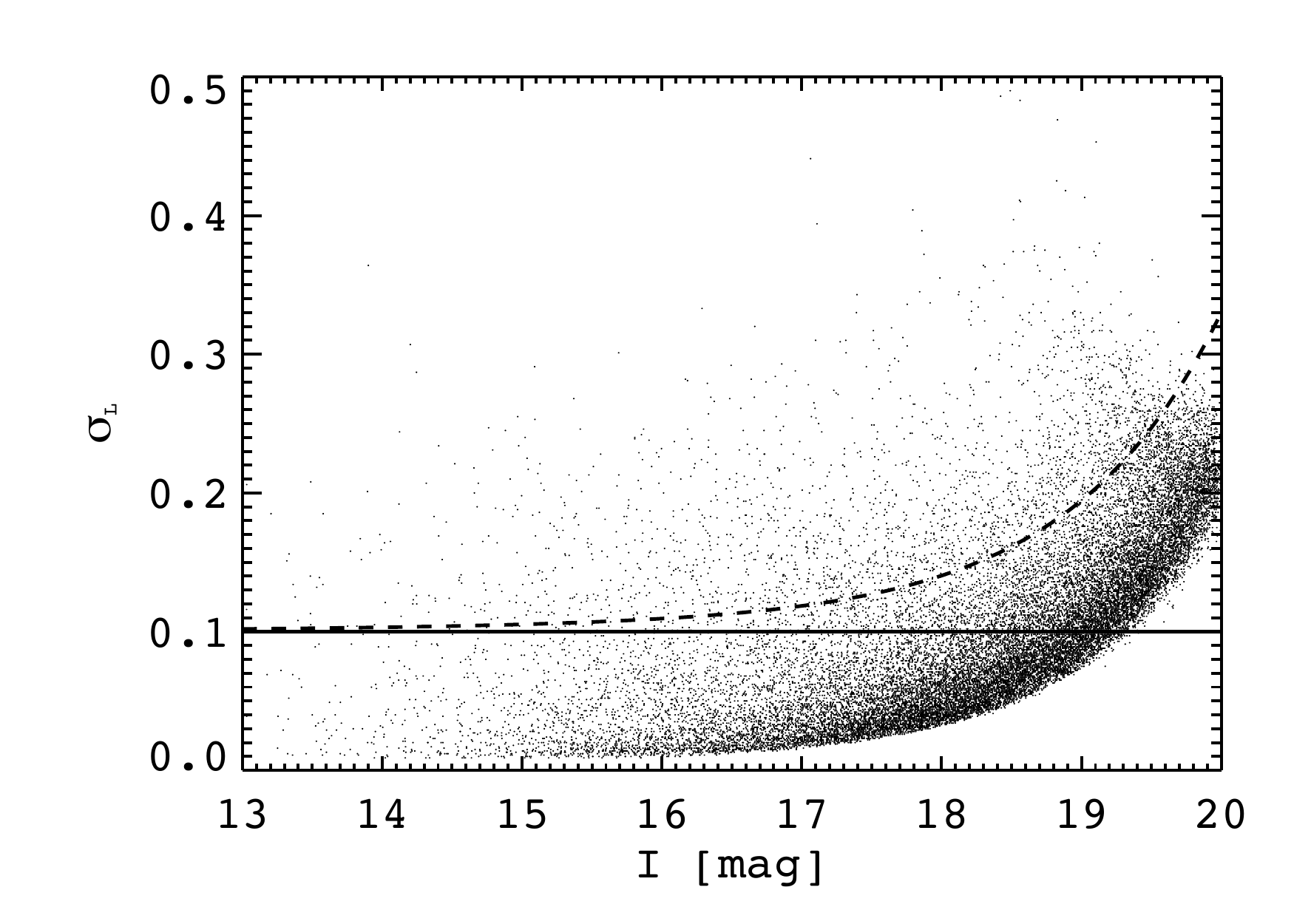}
\caption{The standard deviation in the $I$-band magnitude as a function of the $I$-band magnitude outside of eclipse for the LMC EBs identified by \citet{graczyk11}. To isolate high quality light curves we discarded systems with $I \gtrsim 19.3\,\rm{mag}$ and with a standard deviation in the magnitude distribution outside of eclipse $\sigma_L > 0.1$ mag (the solid line). We also tried to limit the sample by $\sigma_L > \sigma_c + 0.1 $, where $\sigma_c$ is the noise limit as a function of magnitude (dashed line). The choice between $\sigma_L > 0.1$ mag and $\sigma_L > \sigma_c + 0.1 $ DO NOT alter the final number of possible disk candidates inside solid box region in Figure~\ref{fig:ks}.}
\label{fig:magerr}
\end{figure}

We now describe how we reduce the full sample of 26,121 OGLE-III LMC EBs, as defined by \citet{graczyk11}, to ensure reliable kurtosis and skewness measurements. 
Figure~\ref{fig:magerr} shows the computed $\sigma_L$ values as a function of $I$-band magnitude
for the full sample of OGLE-III LMC EBs.
It is clear from this plot that noisy systems tend to be the fainter ones, as one would expect since photometric precision is dependent on brightness. To reduce our sample to systems with high-photometric precision in a simple way, we discarded systems with apparent $I$-band 
magnitudes $m_I \gtrsim 19.3\,\rm{mag}$ and dispersion outside of eclipse $\sigma_L > 0.1$ mag.

If there are too few points within an eclipse window we are unable to accurately measure higher moments of the magnitude distribution.  
Within the eclipse window we counted the number of light curve points,  
$N_{ecl}$. We kept those systems with $N_{ecl}/N \le 0.25$, 
where $N$ is the number of all the observed data points in the light curve.
This criterion also restricted our study to well-detached EBs.
Furthermore, we required that $N_{ecl} > 20$ to ensure reliable kurtosis and skewness measurements of the magnitude distribution.
Lastly, we restricted the sample to deep eclipses. In order to search for 
possible disks around the secondary binary components, 
we chose only systems with primary depth 
$|m(\theta_p) - \mu_L| > 4 \sigma_L$, where $|m(\theta_p)  - \mu_L|$ is an estimate for primary eclipse depth in magnitudes. Independently, during the search for a possible disk around the primary star, we used a similar criterion $ |m(\theta_s) - \mu_L| > 3 \sigma_L$ within the secondary eclipse window. 
We found that less strict criteria introduced a significant number of contaminants into our reduced sample.
Here contaminants are systems that have magnitude distributions within the eclipse window similar
to those hosting discs.

The skewness, $S$, and kurtosis, $K$,  of the magnitude distribution in the primary or secondary eclipse windows 
were computed as
\begin{equation} \label{eq:skew}
 S =  \frac{1}{N} \sum_{i=0}^{N-1} \left(\frac{m_i - \mu}{\sigma_L}\right)^3 \end{equation}
 \begin{equation} \label{eq:kur}
 K =  \frac{1}{N} \sum_{i=0}^{N-1} \left(\frac{m_i - \mu}{\sigma_L}\right)^4 - 3 \end{equation}
where the mean in the eclipse window $\mu = \frac{1}{N} \sum_{i=0}^{N-1} m_i$.
Here $\mu$, the mean magnitude within eclipse,  should not be confused with $\mu_L$, the median magnitude outside of eclipse. 
$N$ is the number of light curve points within the eclipse window in the phase-folded light curve and $m_i$
is the magnitude of each light curve data point.

 \subsection{Skewness and kurtosis moments of the eclipse window magnitude distributions}
 
We first apply our technique to the sample of detached EBs found in the OGLE-III survey 
in the LMC by \citet{graczyk11}.
The OGLE-III survey in the LMC fields covers approximately 40 square degrees and
detected about 32 million LMC sources \citep{udalski08}. Of these,
26,121 have been identified as EBs \citep{graczyk11}.  
The search included all stars brighter than $m_I=20\,\rm{mag}$, with over 120 photometric measurements for each object.
The search for EBs was performed using the method outlined by \citet{graczyk10}
and was restricted to $1.0015 < P< 475\,\rm{days}$.  
Period searches were performed using the phase dispersion minimisation (PDM) method \citep{stellingwerf78}. 

\begin{figure}
\includegraphics[width=\columnwidth, trim= 0.5in 0 0 0 ]{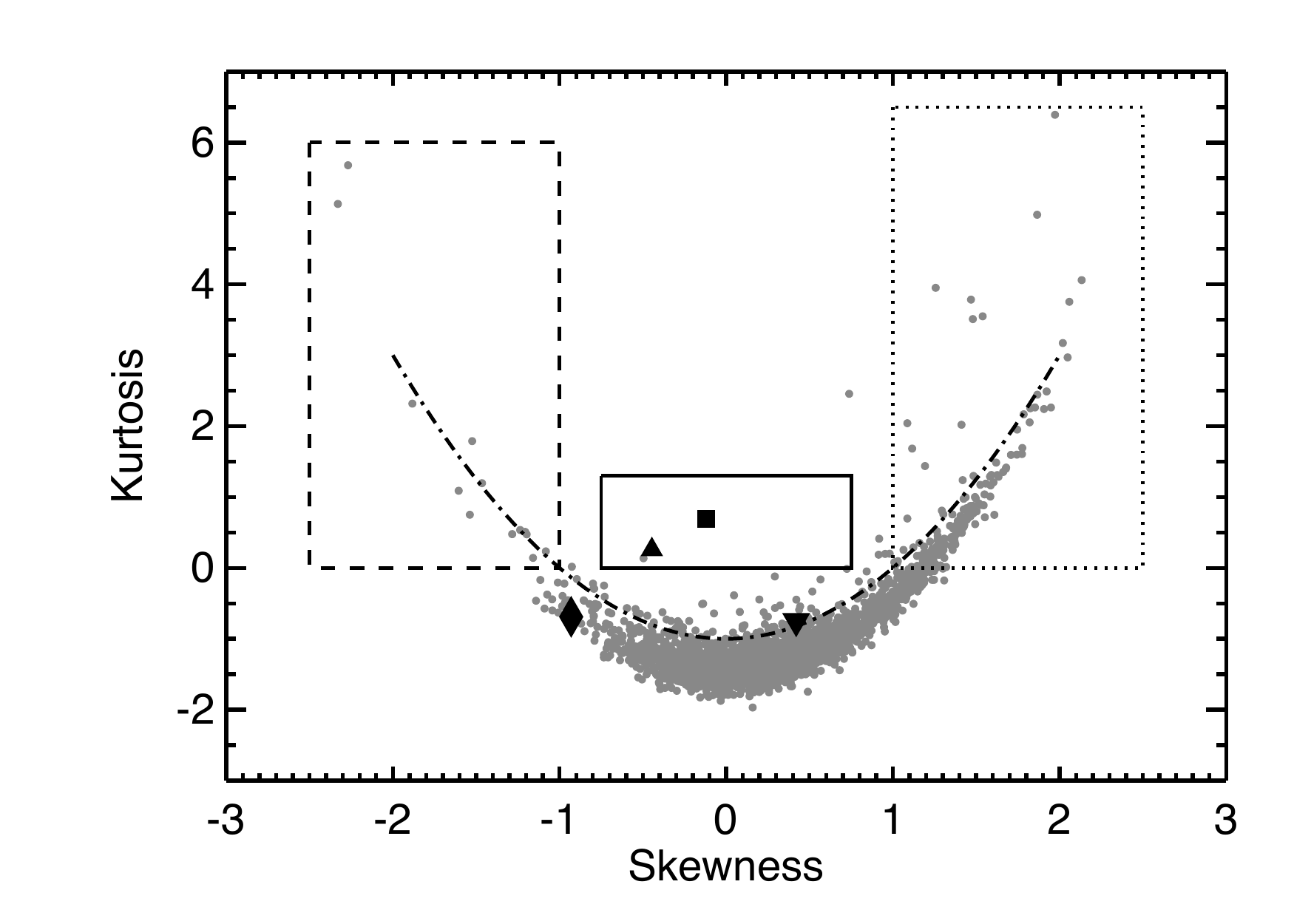}
\caption{The kurtosis vs. skewness computed from the
magnitude distribution within primary eclipse of the sample of 2,823 low-noise LMC EBs. 
OGLE-LMC-ECL-17782 and OGLE-LMC-ECL-11893 are plotted as the square and triangle points, respectively.
We divide the kurtosis vs. skewness plot into four different regions: 
the dashed, solid and dotted line regions, and a region external to these. 
Systems within these regions have different eclipse shape profiles. 
Systems within or nearby the solid line region could be eclipsing disc candidates.
The dash-dot parabolic line ($K=S^2-1$) gives the overall shape of 
the magnitude distribution. 
EE Cep and $\epsilon$ Aur are plotted as filled upside-down triangle and filled diamond, 
respectively.
The skewness and kurtosis measurements of EE Cep and $\epsilon$ Aur do not
distinguish them from other eclipsing binaries.
\label{fig:ks}
}
\end{figure}

Our process of choosing high-cadence, good photometric quality eclipses with large primary depth reduced the LMC sample of 26,121 systems to 2,823 systems.  We refer to this reduced sample as the
low-noise LMC EB sample.
In Figure \ref{fig:ks} we plot skewness vs kurtosis of the magnitude distribution in primary eclipse
for the low-noise subsample of detached OGLE-III LMC EBs.
The two previously known eclipsing disc systems, OGLE-LMC-ECL-17782 and OGLE-LMC-ECL-11893,
 are plotted as filled square  and triangle, respectively.   
They stand out as having low absolute value of skewness  $|S| < 1.0$
 and positive kurtosis $K > 0$. The solid border region in Figure \ref{fig:ks}
is arbitrary chosen for a better view. During the search, we define a parabola ($K = S^2 - 1$, which depicts the property of the ``common'' EB system in LMC) to select those with a significant plateau stage before and/or after the middle of eclipse, and we mean by significant that the plateau has an aggregation of the light curve points excess than the normal distribution ($S = 0$ and $K = 0$). EB systems with in-eclipse $|S| = 1$, has $K = 0$ following the parabolic rules for all the high S/N LMC sample. 
We also compute in the same way the skewness and kurtosis measurements for 
EE Cep (plotted as a filled upside-down triangle) using the V band 
phase-folded data\protect\footnotemark[1]{} \citep{galan12} and $\epsilon$ Aur 
(plotted as a filled diamond) calculated from V band data\protect\footnotemark[2]{} 
between 1982 to 1988.
\footnotetext[1]{\url{http://vizier.cfa.harvard.edu/viz-bin/VizieR?-source=J/A+A/544/A53}}
\footnotetext[2]{\url{http://www.hposoft.com/campaign09.html}}
The skewness and kurtosis measurements of EE Cep and $\epsilon$ Aur do not
distinguish them from other eclipsing binaries, because there is no plateau exhibiting within the eclipse of EE Cep and eps Aur.

\begin{figure}
\includegraphics[width=\columnwidth, trim= 0in 0 0 0 ]{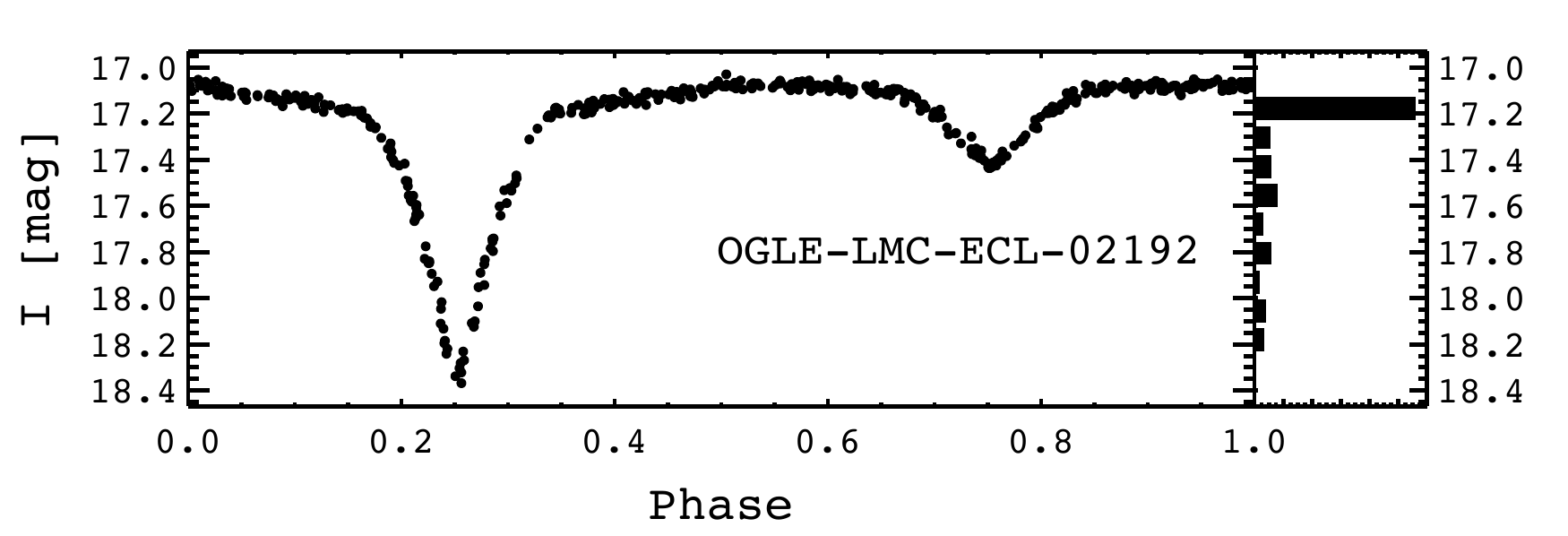}
\caption{The phase-folded light curve of OGLE-LMC-ECL-02192 ($\rm{P} = 1.59\,  \rm{days}$), a detached EB with one component Roche lobe filled. In the right panel is displayed a histogram of the magnitude distribution within eclipse. This is a binary with a positive in-eclipse skewness (it is located in the dotted line region of Figure~\ref{fig:ks}, with $S=1.94$ and $K=4.98$). The ingress and egress of the primary eclipse are at 0.1 and 0.5 phase. The magnitude distribution within eclipse contains a bright tail due to eclipse ingress and egress that gives the distribution a positive skewness.}
\label{fig:02192}
\end{figure}

\begin{figure}
\includegraphics[width=\columnwidth, trim= 0in 0 0 0 ]{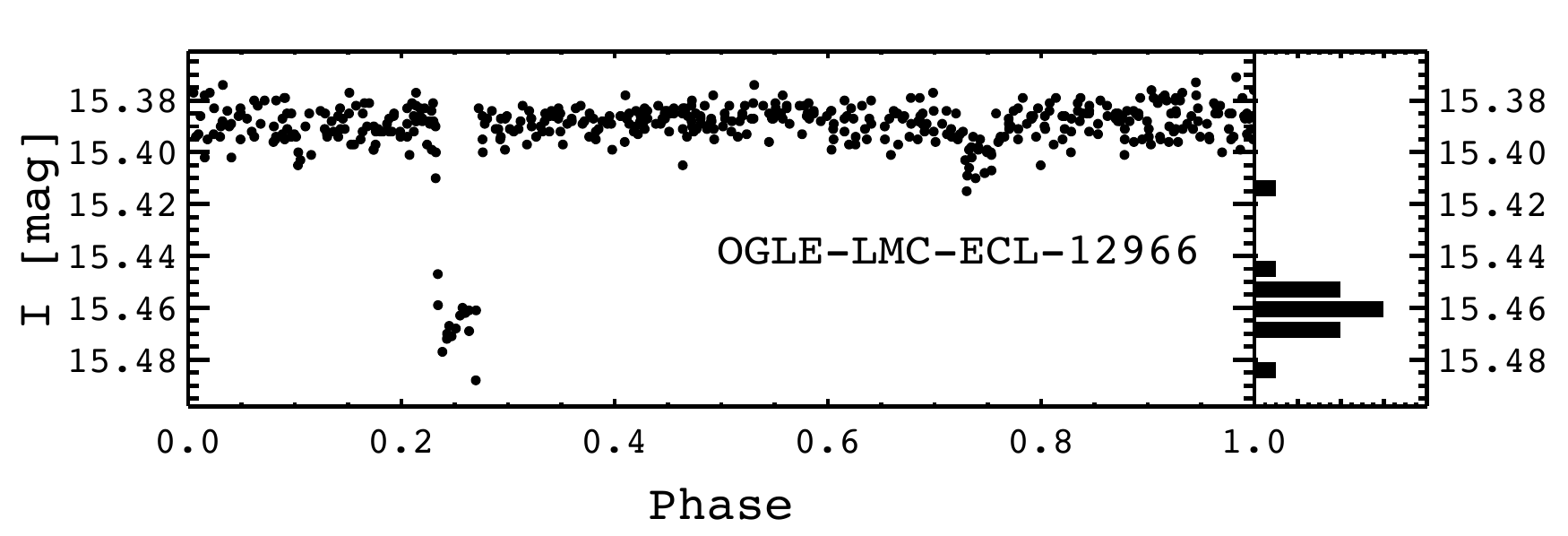}
\caption{Same as Figure~\ref{fig:02192} but for OGLE-LMC-ECL-12966 ($\rm{P} = 239.6\, \rm{days}$), a well-detached EB which has a negative in-eclipse skewness ($S=-1.35$ and $K=1.21$) due to the square shape of the light curve at the bottom  of eclipse. The magnitude histogram (on the right) is narrow with some contribution at fainter magnitudes, 
giving the distribution  a negative skewness  so that it lies 
in the dashed line region in Figure~\ref{fig:ks}.
}
\label{fig:12966}
\end{figure}

\begin{figure}
\includegraphics[width=\columnwidth, trim= 0in 0 0 0 ]{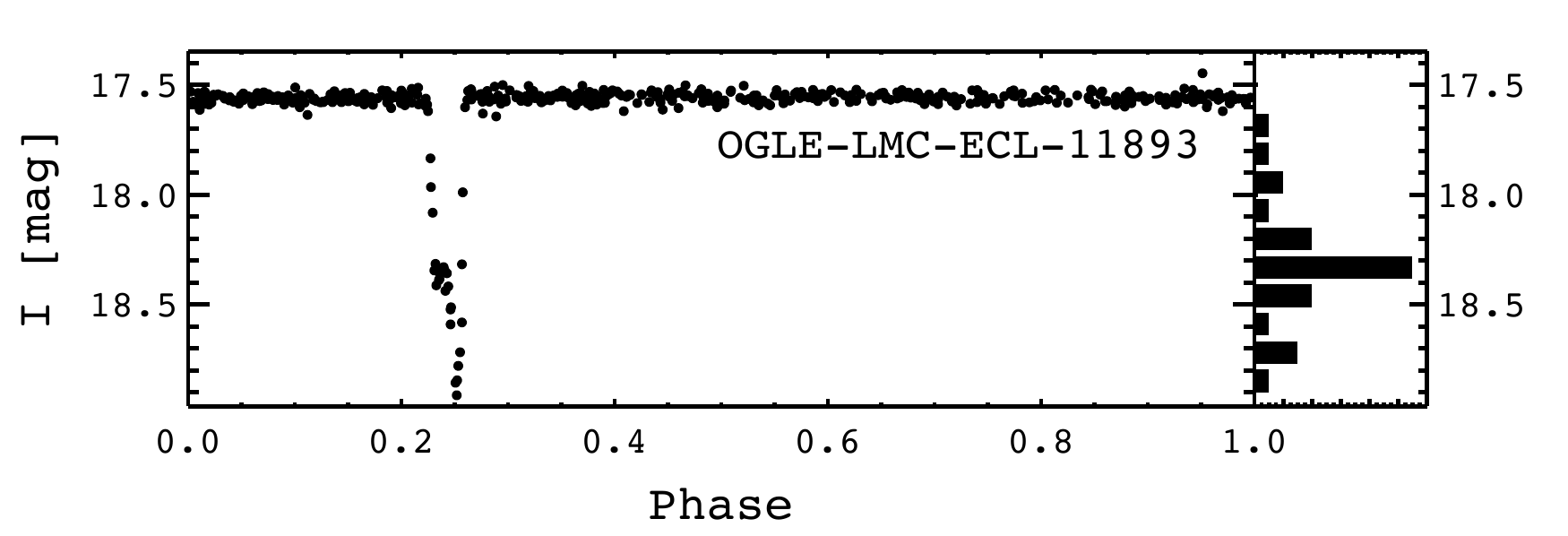}
\caption{Same as Figure~\ref{fig:02192} but for OGLE-LMC-ECL-11893 ($\rm{P} = 468.045\, \rm{days}$), a previously identified eclipsing disc system \citep{dong14}. The magnitude distribution within eclipse has a near zero skewness but a positive kurtosis ($S=-0.45$ and $K=0.25$) due to the platform in the wings of the eclipse.}
\label{fig:11893}
\end{figure}

\begin{figure}
\includegraphics[width=\columnwidth, trim= 0in 0 0 0 ]{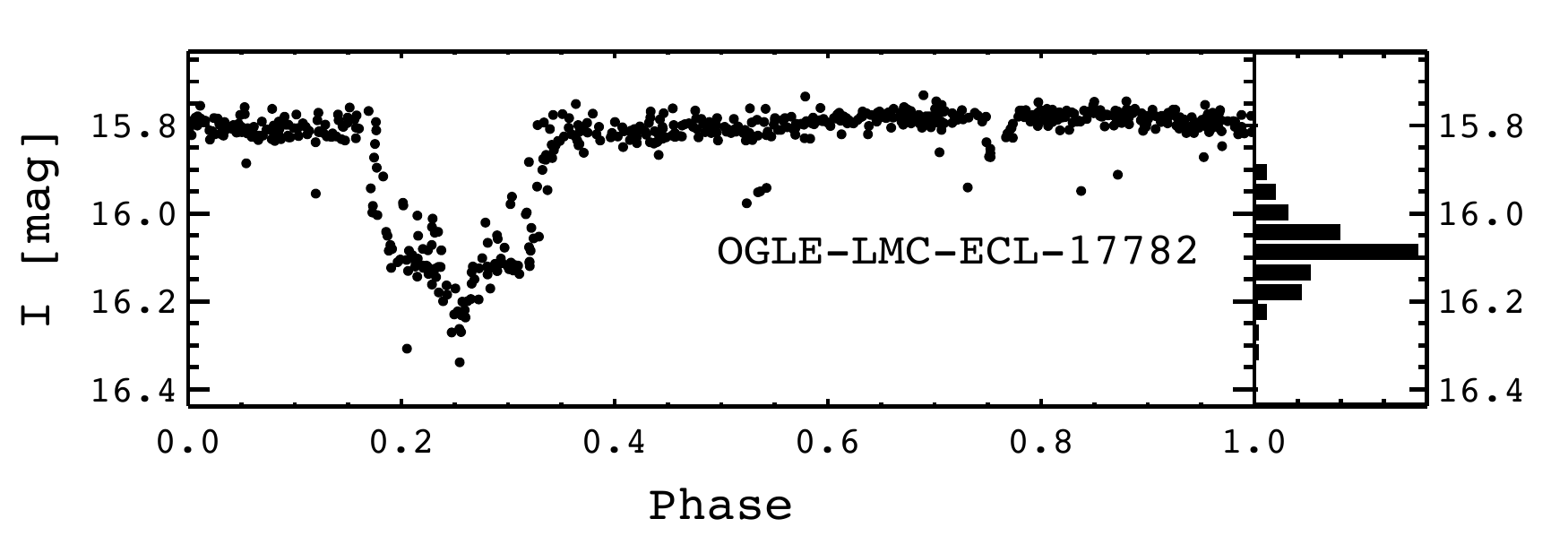}
\caption{Same as Figure~\ref{fig:02192} but for OGLE LMC-ECL-17782 ($\rm{P}=13.353\,\rm{days}$), another previously identified eclipsing disc candidate \citep{graczyk11}). The in-eclipse magnitude distribution of this system has skewness and kurtosis similar to that of OGLE LMC-ECL-11893 ($S=0.12$ and $K=0.89$), which also hosts an eclipsing disc. This eclipse has variable profile and depth.}
\label{fig:17782}
\end{figure}

Why do two of the systems hosting eclipsing discs, OGLE-LMC-ECL-17782 and OGLE-LMC-ECL-11893, stand out in this plot? 
As we can see from Equations~\ref{eq:kur} and \ref{eq:skew}, the kurtosis can be treated as an estimate of the concentration of the magnitude distribution,  whereas the skewness measures the asymmetry of the magnitude distribution i.e. it measures whether the peak is brighter or fainter than the mean in-eclipse magnitude. We divided Figure~\ref{fig:ks} into four regions: a dashed, solid and dotted line region, and a region exterior to these. The dotted line region contains EBs with positive skewness, so light curve points in eclipse  have a peak/aggregation at the beginning and ending of the eclipse -- they tend to have shallow ingresses (see for example the light curve of OGLE-LMC-ECL-02192; Figure~\ref{fig:02192}). In contrast, systems in the dashed line region with negative  skewness, have many light curve points near the maximum eclipse depth -- the eclipses have square bottoms (see the light curve of OGLE-LMC-ECL-12966; Figure~\ref{fig:12966}). When there is a disc around one component of the EB, the light curve may exhibit a ``platform'' at the wings of the eclipse. This is natural because the disc structure is more extended than the central star. Before the secondary occulting the star, the disc occults. If the structure of the disc is not complicated (ringed or lopsided) then there should be a relatively flat stage during the eclipse and this distinguishes OGLE-LMC-ECL-11893 (Figure~\ref{fig:11893}) and OGLE-LMC-ECL-17782 (Figure~\ref{fig:17782}) from the remaining systems in Figure~\ref{fig:ks}.
The skewness and kurtosis are not sensitive to asymmetry in the eclipse profile so EE Cep is not distinguished.
These moments are also not necessarily sensitive to a W-shaped eclipse and so the $\epsilon$ Aur eclipse is
not distinguished.  The skewness and kurtosis do find systems with an unusual number of light-curve data points mid-eclipse. 

\begin{figure}
\includegraphics[width=\columnwidth, trim= 0.1in 0 0 0 ]{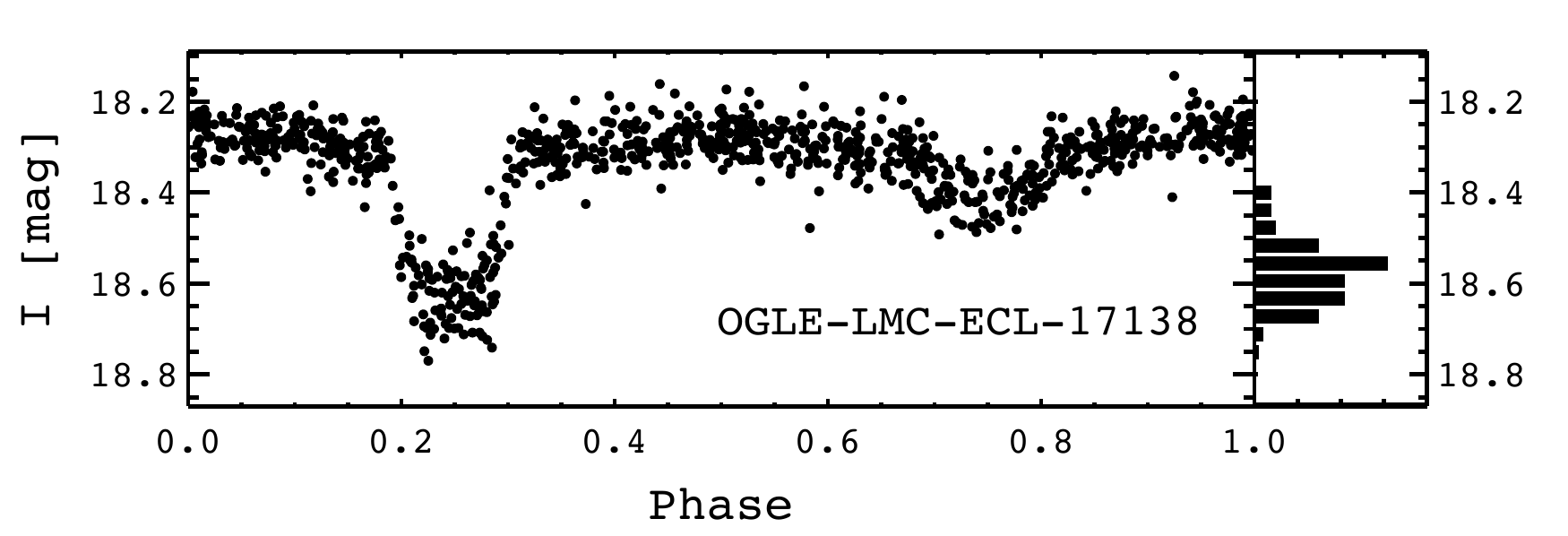}
\caption{Same as Figure~\ref{fig:02192} but for OGLE LMC-ECL-17138 ($\rm{P} = 11.1 \rm{days}$), a contaminant within the solid line region of Figure~\ref{fig:ks} which, according to the histogram of in-eclipse magnitude distribution, has a similar statistical skewness and kurtosis ($S=-0.36$ and $K=-0.02$) as the other two identified eclipsing disc candidates -- OGLE-LMC-ECL-11893 and OGLE-LMC-ECL-17782. 
The odd kurtosis is due to variability in the eclipse depth.}
\label{fig:17138}
\end{figure}

In the solid line region of Figure~\ref{fig:ks} with near zero skewness and positive kurtosis (containing the two systems previously identified with eclipsing discs) we find the system OGLE-LMC-ECL-17138 with the light curve shown in Figure~\ref{fig:17138}.  This system does not host a disc, 
so why are the moments of its eclipse magnitude distribution similar to that of the two eclipsing disc systems? 
We attribute its high kurtosis to variations in eclipse depth.  We found that individual eclipses (seen in the unfolded light curve) were square bottomed and thus it should have a low kurtosis, however because the eclipse depth is variable, the magnitude distribution measured in the phase-folded light curve has a high kurtosis value.  
We can treat this object as a contaminant that must be removed with by-eye inspection of the phase-folded light curve in a search for eclipsing discs. 

\begin{figure}
\includegraphics[width=\columnwidth, trim= 0.5in 0 0 0 ]{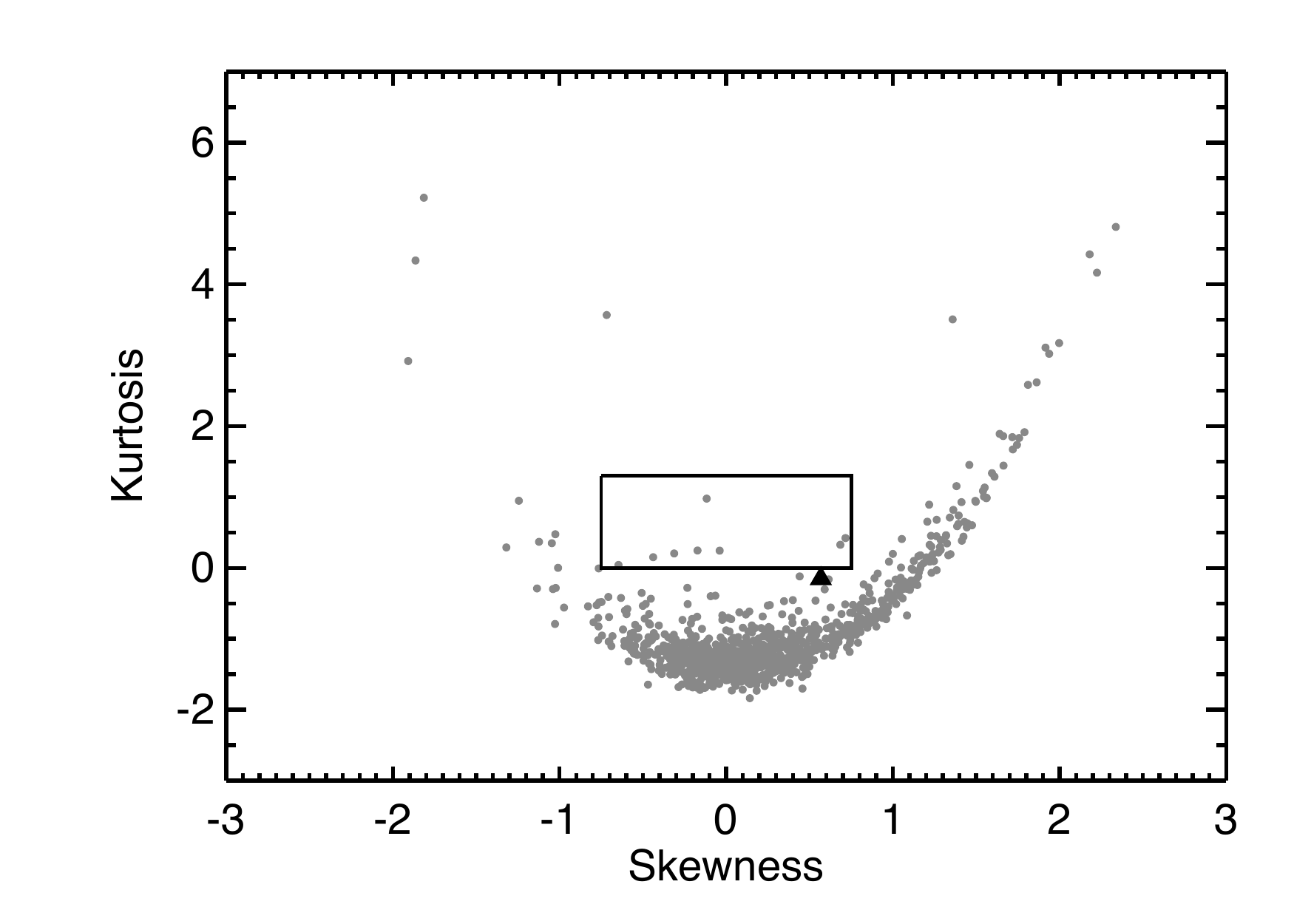}
\caption{Same as Figure~\ref{fig:ks} but for the SMC EBs \citep{Pawlak13}.
We listed those systems within the same solid line region as the low-noise LMC EB sample (see Figure~\ref{fig:ks}), however all of these are rejected as false positives after visual inspection yet for OGLE-SMC-ECL-0007 (plotted in triangle, also see in Figure~\ref{fig:s0007}) outside the solid border shows asymmetric feature inside eclipse.}
\label{fig:smc_ks}
\end{figure}

\begin{figure}
\includegraphics[width=\columnwidth, trim= 0.0in 0 0 0 ]{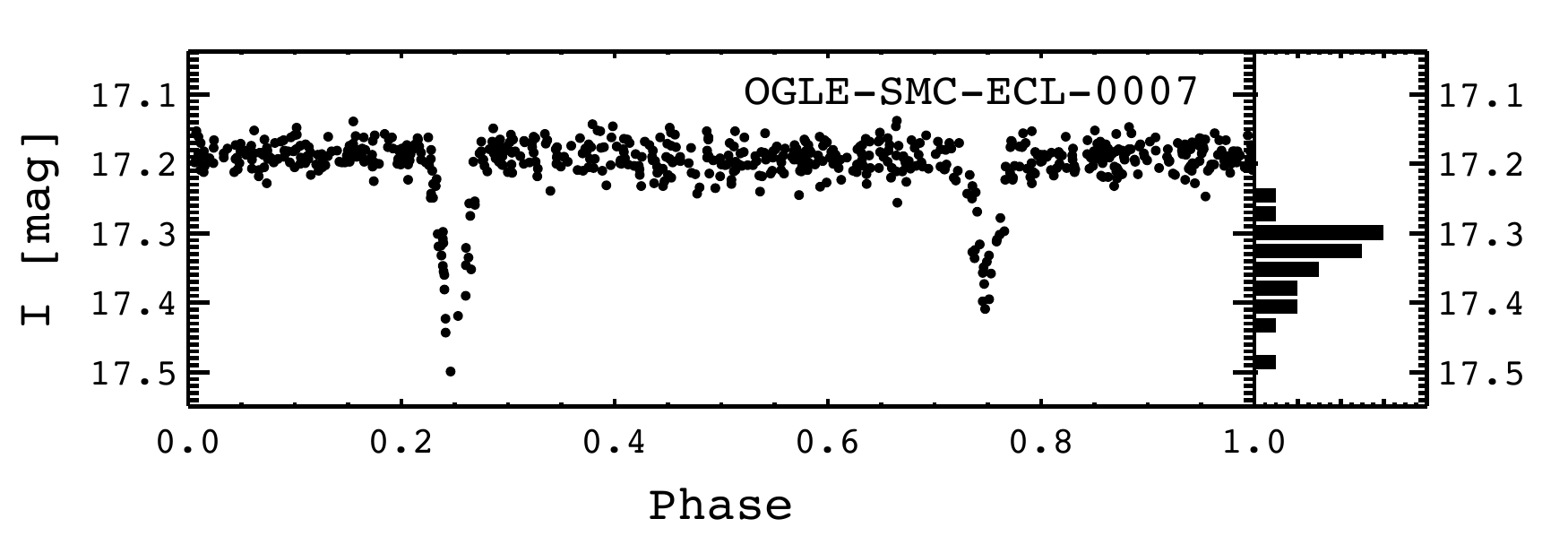}
\caption{Same as Figure~\ref{fig:02192} but for OGLE-SMC-ECL-0007 ($\rm{P}=1.211\,\rm{days}$), an example in the SMC EB sample shows that our method can possibly causing the missing of other interesting systems with  interesting asymmetric features ($S=0.57$ and $K=-0.15$). Since both the primary and secondary eclipse show the same asymmetric shape, and considering the short period of the system, we suspect eclipsing timing variations.}
\label{fig:s0007}
\end{figure}

In addition to analysing the primary eclipses, we also searched i) the secondary eclipses in the low-noise LMC sample, ii) the OGLE-III survey EBs in the SMC \citep{Pawlak13} and iii) the OGLE-III GD EB catalogue \citep{Pietrukowicz13}.  In  the LMC sample, we find low-noise  secondary eclipses in 418 EB systems, selected in the same way as for the primary eclipses.
Of these none were classified as candidate  
disc-hosting objects by our kurtosis-skewness method. 
In the SMC sample, we reduced the EB sample from 6,138 to 748 high S/N systems, and after evaluating the kurtosis and skewness (shown in Figure~\ref{fig:smc_ks}), we found a single system (OGLE-SMC-ECL-0007) 
with a possible disc-like feature (shown in Figure~\ref{fig:s0007}), that
exhibits asymmetric features in both primary and secondary eclipse.
Since this object has a very short period of $1.211\,\rm{days}$, it is unlikely that it hosts two discs. 
Its red color, $V-I = 2.64$ mag \citep{Pawlak13},  and 
short period suggest that it is a foreground system comprised of late-type stars. 
We inspected the unfolded light curves and found that the centre of eclipse varies with respect to the period of the system. Unfortunately, in the unfolded light curves the average number of data points in each eclipse is only approximately 3-4, so it is difficult to make concrete conclusions about the shape of the eclipses. Variations in eclipse shape in the phase-folded light curve could  be explained by eclipsing timing variations
 or variability associated with a stellar wind.

 Lastly, we carry the technique into the GD sample and identified no possible disc features in the GD reduced sample (570 systems). We attribute this to the fact that the typical depth-to-noise ratio is lower than the LMC sample (because of the higher noise level of GD sample than LMC/SMC sample). Moreover, most of the GD stars are old stars, and disc lifetimes are expected to be short.  Notably, the longest EB period in GD sample is only $103.502\,\rm{days}$.  The larger the separation
 between two stars, the larger a disc can fit within the Hill radius of one of the stars.
 Hence we might expect a short period survey would be less likely to discover an eclipsing disc.
  

\section{Properties of the primary stars of the two systems hosting discs in the LMC}
\label{properties_primary_stars}

As shown by \citet{derekas07} and \citet{graczyk11} the EBs in the LMC  
exhibit a bimodality in a colour-magnitude diagram 
with the bluer set near the main-sequence and the redder set likely red
giants and supergiants.  Of the EBs in the OLGE-III LMC sample, how many are near the main-sequence?  
If we define near main-sequence objects as those with $V-I \le 0.5$ mag, then of our
2,823 low-noise EBs we find 2,471 near main-sequence EB systems in the low-noise LMC sample.
Most of the sample is near the main-sequence.


\begin{figure}
\includegraphics[width=\columnwidth, trim= 0.5in 0 0 0 ]{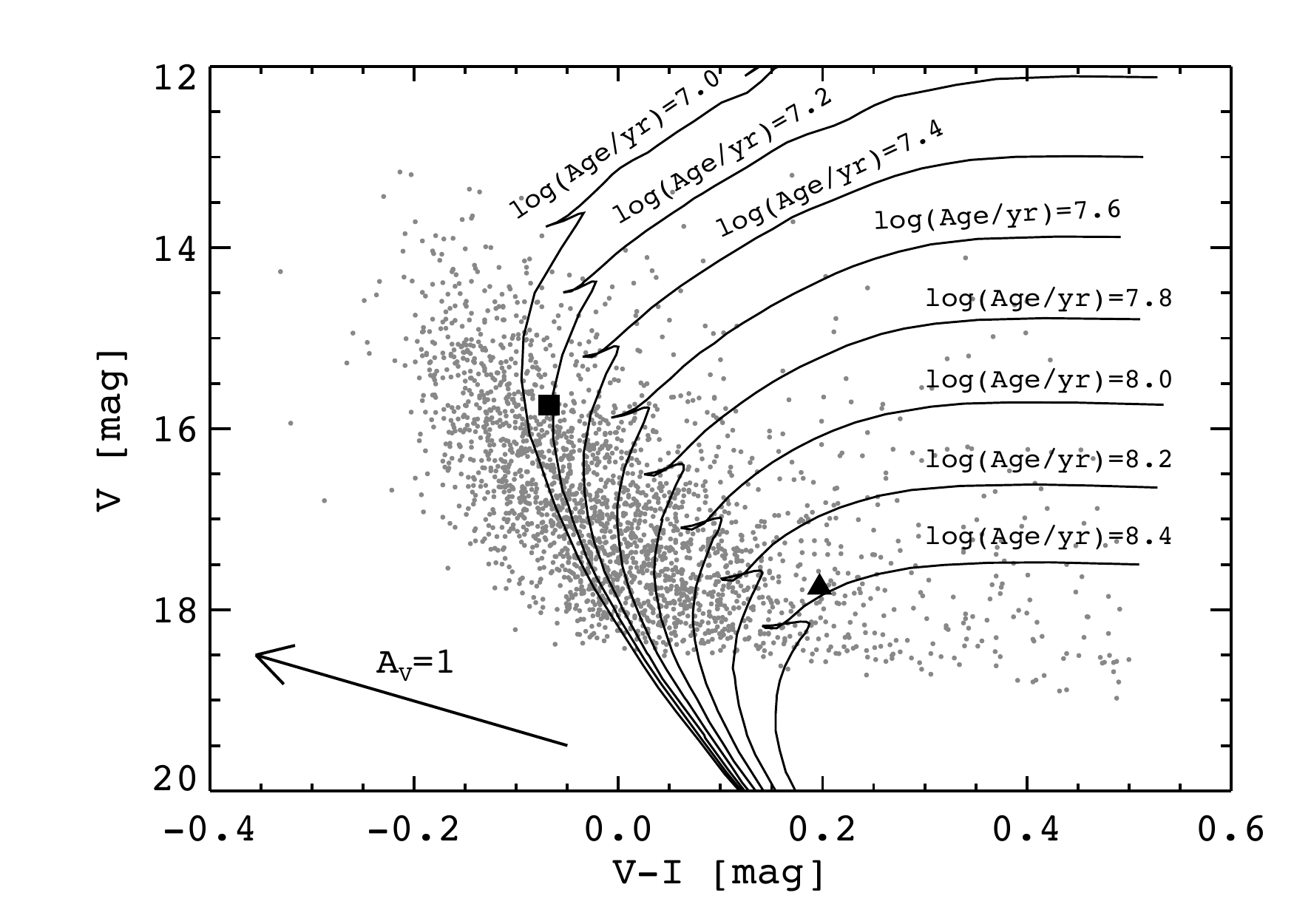}
\caption{Colour-magnitude diagram of the low-noise LMC EB sample but restricted to blue ($V-I \le 0.5$ mag) main-sequence systems. Overlaid are main-sequence isochrones \citep{Marigo08} with ages ranging from $\log_{10}$(age/yr)$=7.0-8.4$ in steps of $\Delta \log_{10}$(age)$=0.2$.  The isochrones have been reddened with $A_V=0.55$ mag and shifted using a distance modulus of $18.48$
 mag.
Also shown is the reddening vector for an extinction of $A_{V}=1\,\rm{mag}$.
The previously identified eclipsing disc systems,
OGLE LMC-ECL-17782 (filled square) and OGLE-LMC-ECL-11893 (filled triangle) are plotted using
photometry by \citet{graczyk11}, and
are consistent with being  near the main-sequence.   
%
}
\label{fig:cmd}
\end{figure}

Figure~\ref{fig:cmd} shows a colour-magnitude diagram of the bluer systems from the low-noise LMC sample. As we can see,  OGLE-LMC-ECL-17782 (filled square) and OGLE-LMC-ECL-11893 (filled triangle) have colours similar to stars
near the main-sequence. 
 %
We used 
the Padova internet server\footnote{\url{http://stev.oapd.inaf.it/cgi-bin/cmd}} to create a grid of \citet{Marigo08} isochrones with a metallicity of $Z=0.006$ (typical value for the LMC) spanning the age range $\log_{10}$(age/yr)$=6.0-9.0$ with a constant age step of $\Delta \log_{10}$(age)$=0.1$ in the OGLE photometric system. We leave the photometric measurements of the EB objects in the apparent magnitude-colour plane and instead transform the theoretical isochrones. For this, we adopt the average extinction towards the LMC of $A_V=0.55$ mag \citep{zaritsky04} and the distance modulus $dm=18.48$ mag \citep{Walker12}. 
Figure~\ref{fig:cmd} shows these isochrones overlaid on the observed colour-magnitude diagram.

\subsection{OGLE-LMC-ECL-17782}

 \begin{table}
 \caption[]{Broadband photometry for OGLE-LMC-ECL-17782.}
 \centering
 \begin{tabular}{c l c l c}
\hline
Band  & mag  & Reference & mag & Reference \\
\hline
$U$  & $15.356\pm 0.027$  & Z04 & 14.38 & M02\\
$B$ & $15.563\pm 0.062$ & Z04 & 15.13 & M02\\
$V$ & $15.711\pm 0.027$ & Z04 & 15.784 & U12\\
$V$ & 15.793                    & D07 & 15.21 & M02 \\
$V$ & 15.737                    & G11 &  15.793 & D07      \\
$R$ & 15.888                   & D07    &          &         \\
$I$ & $15.855\pm 0.035$ & Z04 & 15.895 & U12 \\
$I$ &   15.805                   & G11 &              & \\
$J$  & $15.97 \pm 0.02$   &   K07 & $15.913\pm 0.039$ & C06 \\
$H$  & $15.98\pm 0.02$    &   K07  &$15.916\pm 0.074$ &C06 \\
$K_{_{\rm{S}}}$ & $15.92\pm 0.07 $  & K07 & $15.907\pm 0.137$ & C06 \\
$\left[3.6\right]$ &  $ 15.951 \pm  0.076$ & M06 & &\\
$\left[4.5\right]$ & $15.853  \pm   0.088$ & M06 && \\
\hline
\end{tabular}
\medskip \\
Sources used to compile the literature photometry are as follows.
Z04 -- \cite{zaritsky04};
K07 -- \cite{kato07};
M06 --  \cite{meixner06};
C06 --  \cite{cutri12};
U12 --  \cite{ulaczyk12};
M02 --  \cite{massey02};
G11 -- \cite{graczyk11};
D07 --  \cite{derekas07}.
\end{table}

Photometry  of OGLE-LMC-ECL-17782 compiled from recent photometric surveys is listed in Table 1. 
Unfortunately the $UBV$ magnitudes and colours
reported by \citet{zaritsky04} and \citet{massey02} are
not consistent, making it difficult to simultaneously constrain the age, mass and the extinction.  
We consider the possibility that the discrepancy could be due to some observations taken during eclipse.
The photometric measurements by \citet{massey02} were taken on five nights, UT 1999 Jan. 8, UT 2001 Mar. 28-30 and UT 2001
Apr. 1.  Using the epoch of primary eclipse minimum Heliocentric Julian Date (HJD) of  2453563.2912 and 
a period of 13.352899 days \citep{graczyk11}, we estimate that an eclipse could have taken place during  2001 April 01 at noon UT
and on 1999 January 07 at 23 hours UT.  The observations by \citet{massey02} 
were taken at CTIO which has local time a few hours earlier than UT.
Since the eclipse is broad,  either during 1999 Jan 8 or 2001 April 1 the system could have been in eclipse.  
If the system were in eclipse during some of the observations, then these would be fainter than the others.
However Massey's observations are brighter than those of \citet{zaritsky04} which are consistent with the
measurements by \citet{derekas07,ulaczyk12}.  Consequently observations during eclipse cannot account for
the discrepancy.  One possibility is that the disk itself shields the secondary making the entire system dimmer.
During epochs with no disk, the system could have been brighter and
Massey's observations could have been taken outside of eclipse but during a time when the secondary was not shielded by
a disk.   The OGLE III observations do not show evidence of brightness variations outside of eclipse \citep{graczyk11}, 
but rare epochs of brightening can be exhibited by B stars \citep{rivinius13}.
The observations by \citet{zaritsky04} were taken between 1995 October and 1999 December, but they did not specify
dates for individual fields.  

While the $UBV$ colors are uncertain, there is agreement in the $V$ and $I$ magnitudes. 
As the estimated primary and secondary stellar radii affect the length of the eclipses, we simultaneously searched
for primary and secondary stellar parameters consistent with the $V,I$ photometry and the shape of the light curve.

The primary and secondary eclipses are separated by 0.5 in phase so the orbit is circular.
The deepest parts of both the primary and the secondary eclipse are triangular shaped and this implies that
the radii of the primary and secondary star are similar.  However, triangular shaped eclipses with rounded bottom
can also arise if the secondary
 has a larger radius than the primary (perhaps as big as a factor of 2) and the orbit is slightly tilted (is a few
 degrees from oriented edge-on to the viewer). 

 To improve on estimates for the nature of the  stars, we roughly modeled the I band light curve with the program
 NIGHTFALL.\footnote{
 For more details, see the Nightfall User Manual by Wichmann (1998) available at the URL: \url{http://www.hs.uni-hamburg.de/DE/Ins/ Per/Wichmann/Nightfall.html}.  The program takes into account limb darkening, gravitational darkening/brightening, and the shape of the stellar equipotential surfaces.}
 We adopted a model with three components, primary and secondary stars and a disk, which are black bodies,
  and by trial and error searched through
 parameter space for a reasonable fit to the $I$-band light curve.  
 We assume that the disk extends to the Roche lobe of the secondary star.
 We find that the secondary must nearly be as massive as  the primary star, $M_2 \gtrsim 0.8 M_1$, 
 otherwise the secondary eclipse length (taking into account the broad wings), 
 due to the disk that fills the Roche lobe, 
 could not be as wide as observed. 

\begin{figure}
\includegraphics[width=\columnwidth, trim= 0.2in 0 0 0 ]{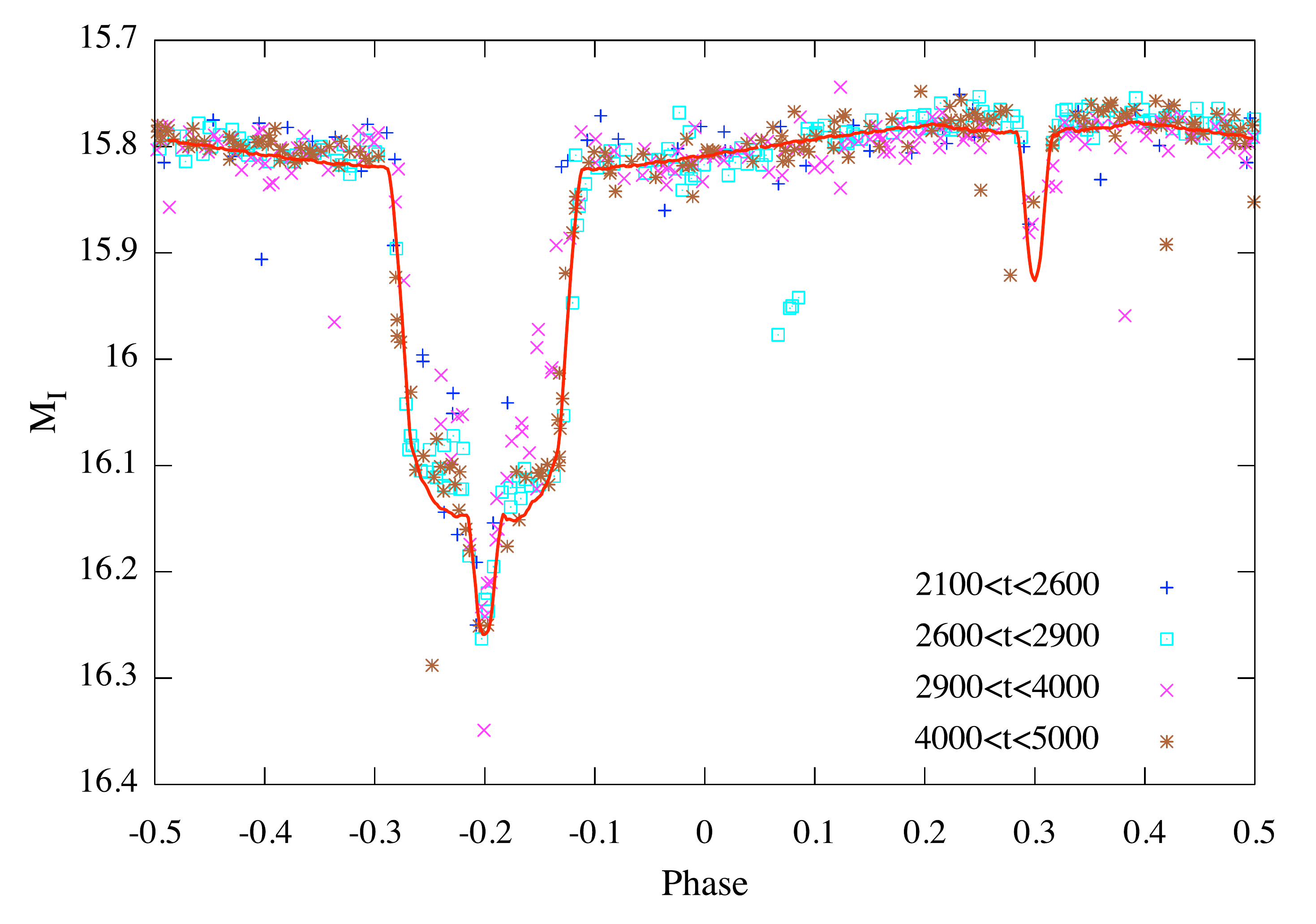}
\caption{Modeled light curve (red line) with folded I band OGLE III light curve (points) for  OGLE-LMC-ECL-17782
(from \citealt{graczyk11}).
The different types of points correspond to different epochs.   The time ranges for the different
points are shown in the key with $t=HJD-2450000$.  Parameters for the NIGHTFALL model are listed in 
Table \ref{tab:nightfall}. To note that we manually shift the primary minimum to $\theta_p=-0.2$ during the fit.
}
\label{fig:nightfall}
\end{figure}

 The NIGHTFALL lightcurve model in I-band, is shown in Figure \ref{fig:nightfall} and parameters
 listed in Table 2.  
 The modeling software was not capable of adjusting disk opacity, assuming that the disk
 emits like a black body, but we were able
 to match the disk eclipse depth by adjusting the disk thickness and orbital inclination.   
 Using interpolated values from the isochrones,
 we adjusted the model so that the predicted $V$ magnitude (summing fluxes from both stars)
 is approximately that observed, the predicted colors
 are similar to those observed, and
  the effective temperatures and luminosities of the two stars lie on a single isochrone by \citet{Marigo08}.
 Along with an approximate fit to the light curve, these combined constraints give a estimated
 value for the masses of the primary and secondary stars, their effective temperatures and extinction
 to the object. 
 The best model we found has a primary with effective temperature of 29000K and luminosity of
$1.4 \times 10^4 L_\odot$ consistent with a mass of 14.3 $M_\odot$,  an age of 5 million years and 
consistent with an extinction of $A_V \sim 0.4$ mag.
Taking into account light from both primary and secondary stars this model
has an absolute magnitude of $M_V = -3.1$ mag (corrected for extinction).
The predicted colors (from the sum of light from both stars)
are $U-B = -0.92$ mag, and $V-I = -0.12$ mag. 
The $V-I$ color is approximately
consistent with the observed $V-I =-0.111$ mag  by \citet{ulaczyk12} and $V-I=-0.144$ mag by \citet{zaritsky04}.
This estimated extinction to the object itself is somewhat lower than the mean value estimated over a large region  
of $A_V = 0.55$ mag  \citep{zaritsky04}. 
If the inferred extinction were significantly higher, then the reddened main sequence is too red to match the observed $V-I$ colour.
If the extinction were lower ($A_V \lesssim 0.3$ mag) then the primary star must be older
and the estimated radius of the primary star
is larger (to match the observed luminosity), giving modeled eclipse lengths exceeding those observed. 

 The disc is associated with the outer shallower part of the primary eclipse.
 A shallow disk related feature is also seen in the model in the wings of the secondary eclipse.
 The software does not take into account reflection off the disk but it is likely that this accounts
 for the shallow brightening outside of eclipse with a maximum at secondary eclipse.
 We have added into the model a small sinusoidal variation in amplitude of 0.025 mag that peaks at the
 phase of the secondary eclipse.
 The software does take into account reflections off the secondary, and we found by modeling a system
 with a secondary star the same  area and temperature as the disk, that 
 a disk albedo of 0.2 would account for this small amplitude.
A prediction of this model is that when there is no disk, the modulation would no longer be present.
However, the secondary could be partly obscured by the disk or be a compact object. 
The region on the sky has recently been observed in X-rays with the XMM satellite, however
 there is no X-ray source at the position of the object.

%
 
  \begin{table}
 \caption[]{Light Curve model for  OGLE-LMC-ECL-17782 \label{tab:nightfall}}
 \centering
 \begin{tabular}{ll}
\hline
Orbital Period & 13.3525 days \\
Primary $T_{eff}$  & 29000  K  \\
Secondary $T_{eff}$  & 25500 K   \\
Disk $T_{eff}$  & 6000 K  \\
Primary Roche Fill & 0.176\\
Secondary Roche Fill & 0.158\\
Disk  Roche Fill & 0.97 \\
Orbit Inclination & 87.5$^\circ$ \\
Orbital eccentricity & 0.0 \\
Disk aspect ratio $H/R$  & 0.03 \\
Mass ratio $M_2/M_1$ & 0.8 \\
Primary mass $M_1$     &  $14.3 M_\odot$ \\
Total mass  & $26 M_\odot$ \\
Primary Radius  & $4.6 R_\odot$ \\
Secondary Radius & $3.7 R_\odot$ \\
Disk Radius  & 32.6 $R_\odot$, 0.15 AU \\
Semi-major axis & 70.0 $R_\odot$, 0.33 AU \\
Primary Luminosity & $1.3 \times 10^4 L_\odot $\\ 
Secondary Luminosity & $5000 L_\odot $\\ 
\hline
\end{tabular}
\medskip \\
The NIGHTFALL model for the light curve includes three components, a primary star, a secondary star and a disk.
The model  is shown with the folded light curve by \citet{graczyk11} in Figure \ref{fig:nightfall}.
\end{table}

If the disk completely covers the primary during eclipse, its 
 opacity would be $\tau \sim 0.3$ corresponding to the
  drop in magnitude of approximately $0.3\,\rm{mag}$ in the $I$-band.
  If the disk only partially covers the primary, then its opacity could be higher.
  Based on the light curve model  we estimate a disc radius  $\sim 0.15\,\rm{AU}$ (based on
  a total mass of 26$M_\odot$).
 If the primary star has luminosity $\sim  1.3\times 10^4\,\rm{L_\odot}$ then the temperature of an irradiated disc
 that absorbs 30 per cent of the primary starlight at a radius of $0.33\,\rm{AU}$
 would be at least $\sim 5500\,\rm{K}$.  It should be hotter taking into account 
 the radiation from the secondary.    
The large fraction of the orbit 
 spent in eclipse suggests that the disc would intersect approximately a few per cent of the light. 
  As is true for some Be stars (e.g. \citealt{waters88,dougherty94}), the disk could be responsible for an infrared excess. 
  A disc temperature of $5500\,\rm{K}$ is above a dust sublimation temperature.
 Even though the eclipse light curve resembles that of a diffuse dusty disc surrounding a star,
 the disc opacity cannot be due to dust.  Assuming that the disc
 it covers a large fraction of the primary during eclipse, the disk opacity is more likely to be due to 
 Thompson scattering and bound free transitions (photo-ionization of HI, see section 2 by \citealt{rivinius13}) rather than
 free-free absorption (based on the disc size scale, estimated temperature and eclipse depth). 
 The light curve model is not well constrained as we lack spectroscopic measurements. 
 Spectroscopic measurements would measure radial velocities and search for emission lines
 and are needed to
  place better constraints on the nature and spectral type of the two stars and on the nature of the disk.
 
 
 
 \subsection{OGLE-LMC-ECL-11893}
 
 As discussed by Scott et al. (in prep.),  an optical spectrum of OGLE-LMC-ECL-11893 
 gives a spectral type of B9{\footnotesize III}.   The
spectrum implies that the primary star has evolved off of the main-sequence, has somewhat higher
extinction than the mean value for the LMC and an age of approximately $150\,\rm{Myr}$.
The spectral type makes it possible to estimate the extinction and hence the absolute $V$-band magnitude.
For a B9 spectral type, the star is too bright to lie exactly on the main-sequence.
 Even though the period of this system is $468\,\rm{days}$, because  the primary star is luminous,
the disc temperature could be near the dust sublimation temperature.
This object is discussed in more detail by Scott et al. (in prep.) along with models for the eclipse profile.


\section{Speculation about the fraction and types of objects that host eclipsing discs}
\label{speculation_fraction_discs}

Compared to previous surveys such as OGLE-II \citep{Wyrzykowski03}, the OGLE-III LMC survey has a much longer observing time span (about two times that of the OGLE-II survey) in addition to better photometric precision and therefore the detection rate for eclipsing binaries in OGLE-III is twice that of the OGLE-II survey \citep{graczyk11}.  The completeness of the LMC EB catalog is $\sim 90$ per cent (\citealt{graczyk11}, and it could even have higher completeness for stars with apparent $I$-band magnitudes brighter than 18). Because of this high level of completeness we digress here on the probability of detecting a disc in the LMC EB population.

The fraction of young stars in the Galaxy with detected discs decays as a function of
age \citep{haisch01}.
The disc lifetime is shorter in early-type stellar systems compared to late type systems, with only a few
percent of A-F stars exhibiting  discs at
$3\,\rm{Myr}$ compared to $30-35$ per cent of T-Tauri stars \citep{hernandez07}.
The fraction of young stars hosting discs also depends on multiplicity.
In the $1-2\,\rm{Myr}$ old Taurus-Auriga star-forming region, \citet{harris12} find that 
only one third of stars in multiple-star systems harbour dusty
discs that emit  detectable ($\gtrsim 10\,\rm{mJy}$) millimetre radiation and
this is to be compared with two thirds of detectable discs in single-star systems.
Putting together these estimates, we might conservatively estimate
that a few percent of stars in a binary system comprised of early-type stars might
host a disc in a sample comprised of stars with  an age less than $1\,\rm{Myr}$, 
and 1/10th of this number would host a disk if the sample contained stars up to ages of $10\,\rm{Myr}$.

Because of the short lifetime for circumstellar discs we expected to primarily discover eclipsing
discs in young main-sequence objects, though a disc could also be formed via mass transfer.   
Of the  2,471 early type EB systems in the low-noise LMC sample, there
are greater than or equal to two objects hosting discs.   Two are discussed here, however
additional disks may be found with a search for asymmetric or W shaped eclipses that would
find objects similar to EE Cep and $\epsilon$ Aur. 
This fraction, $\gtrsim 1/1000$,  is not obviously inconsistent with the fraction of objects
hosting primordial discs that we might have expected to find in this number of EBs.
However, neither of the two discs discovered are consistent with a primordial circumstellar disc.
OGLE-LMC-ECL-11893, with an estimated age of 150 Myr \citep{scott14} is too old and 
OGLE-LMC-ECL-17782 is not a dusty disc, but a transient one, and so more likely to be recently formed or formed
by a stellar wind \citep{graczyk11} or a decretion disk about a rapidly rotating star \citep{rivinius13}.
As dust sublimates above a temperature $\sim 1400\,\rm{K}$, and disc lifetime depends on radius,
we would have expected to primarily find discs in widely separated binaries,  however OGLE-LMC-ECL-17782
has only a $13.3\,\rm{day}$ period.

Last but not least, our detection rate of possible disc-features in a complete EB sample can be underestimated because our kurt-skew method does NOT distinguish the two previously-known disc-existing systems (EE Cep \& eps Aur) with most of the LMC EB samples. That makes the detectability of disc candidates within EB systems drop to 50\%. Since Neither EE Cep nor eps Aur  show plateau in the light curve before and/or after the center of eclipse, our way of finding other systems with possible disc-like features, however as a complement, gives an complete searching for systems like OGLE-LMC-ECL-17782 or OGLE-LMC-ECL-11893. Changing the border line in Figure~\ref{fig:ks} could include systems with asymmetric feature like EE Cep or W-shaped feature like eps Aur, but we recommend developing other methods of detecting those features. 

\section{Summary and Conclusion}
\label{conclusions}

Following the discovery (by \citealt{graczyk11}; \citealt{dong14}) 
of two EB systems hosting eclipsing discs in the OGLE-III LMC
survey identified by \citet{graczyk11}, we searched for an automated way to
find systems hosting eclipsing discs from the light curves of previously identified EBs.
We are searching for light curves with transits that are asymmetric, W-shaped or have extended  platforms external to 
a primary or secondary stellar transit.  
The technique of using statistical moments of light curve magnitude distributions to automatically 
classify variable stars \citep{graczyk10} 
can be adapted to search for eclipsing discs by using the magnitude distributions within eclipse.
We find that the two previously identified eclipsing disc systems in the LMC stand out 
as having positive kurtosis and a low absolute value of skewness in
the magnitude distribution of light curve points within primary eclipse.
Additional objects have similar moments in their eclipse
magnitude distributions but do not display  disc-like absorption profiles.  
We find that these contaminants are systems which show variable eclipse depths or transit timing variations.

We have applied this search technique to identify eclipsing disc systems in the low-noise LMC EB sample (both primary and secondary eclipse), the SMC EBs identified by \citet{Pawlak13} and the GD EBs catalogued by \citet{Pietrukowicz13}, however
 we have failed to find any new disk candidates.
 
 The moments of the magnitude distribution in eclipse reveal light curves with an excess of data points
at mid-eclipse brightness, however they do not reveal asymmetric eclipses such as those in the 
EE Cep and $\epsilon$ Aur  disk systems.   

The sample of EBs identified by the OGLE-III LMC survey 
has been estimated to be $\sim 90$ per cent complete \citep{graczyk11}.    
Here we determine that 2 of the low-noise 2,471 main-sequence EBs host a  disc that is seen in eclipse.
We would have expected to find discs around young stars (as disc lifetime is short)
and  in widely separated binary systems, as disc temperatures would exceed the dust sublimation temperature
near the star, disc lifetime depends on disc radius
and there is more room within a  Roche radius to host a disc in a binary when the orbital separation is large.
Contrary to what we expected, OGLE-LMC-ECL-11893 is $\sim 150\,\rm{Myr}$ old,
and OGLE-LMC-ECL-17782 hosts a transient but compact ($\sim 0.1\,\rm{AU}$) hot ($\sim 6000\,\rm{K}$) disc 
in a system with a  B star.  
Neither disc is consistent with being a primordial circumstellar disc.
Perhaps longer and deeper surveys are needed to find eclipsing counter parts to 
circumstellar discs seen in the infrared and millimetre in Galactic studies.

As photometric precision, cadence and coverage improves we hope that additional disc systems can be discovered, 
thereby providing
tighter constraints on the disc
fraction in young stars and allowing brighter targets to be discovered and subsequently studied.
The nature of OGLE-LMC-ECL-17782 will be better understood with spectroscopic observations.

\subsection*{Acknowledgements}
We are grateful to the OGLE survey for providing light curves of the EBs. 
We especially thank Subo Dong for bringing to our attention the OGLE-LMC-ECL-11893 system 
discovered in the OGLE-III LMC EB sample.
This work was in part supported by NASA grant NNX13AI27G, National Basic Research Program of China (2013CB834900), 
National Natural Science Foundations of China(Nos.10925313, 11333002), The Strategic Priority Research Program-The Emergence of Cosmological Structures of the Chinese Academy of Sciences (Grant No. XDB09000000), the Natural Science Foundation for the Youth of Jiangsu Province (NO. BK20130547), National Natural Science Funds for Young Scholar(No. 11003010) and a research mobility
travel grant from the University of Rochester as part of the Worldwide Universities Network.
We thank Nanjing University for their gracious hospitality during June 2013.
We thank Josh Pepper for helpful discussion. We thank Ningxiao Zhang for confirming the non-existence of 
OGLE-LMC-ECL-17782's X-Ray detection in XMM. 


\begin{thebibliography}{}

\bibitem[Bell et al.(2013)]{bell13}   
Bell, C. P. M., Naylor, T., Mayne, N. J., Jeffries, R. D., \& Littlefair, S. P.	
2013, MNRAS, 434, 806	

\bibitem[Bouwman et al.(2006)]{bouwman06}
Bouwman, J., Lawson, W. A., Dominik, C., Feigelson, E. D., Henning,
Th., Tielens, A. G. G. M., Waters, L. B. F. M. 2006, ApJ, 653, L57

\bibitem[Bressan et al.(2012)]{bressan12}
Bressan, A., Marigo, P., Girardi, L., Salasnich, B., Dal Cero, C.,  Rubele, S., Nanni, A. 2012,
MNRAS, 427,  127

\bibitem[Chadima et al.(2011)]{chadima11}
Chadima, P., Harmanec, P., Bennett, P. D., et al.\ 2011, A\&A, 530A, 146


     
\bibitem[Cutri et al.(2012)]{cutri12}
Cutri, R.~M.,  Skrutskie, M.~F.,  van Dyk, S., et al. 2012, 2281
  2MASS 6X Point Source Working Database / Catalog (Cutri+ 2006),
VizieR Online Data Catalog: 2MASS 6X Point Source Working Database

\bibitem[Derekas et al.(2007)]{derekas07} 
Derekas, A., Kiss,  L.~L., \& Bedding, T.~R.\ 2007, \apj, 663, 249 

\bibitem[De Rosa et al.(2013)]{derosa13}
De Rosa, R. J., Patience, J., Wilson, P. A., et al. 2013, MNRAS, 437, 1216

\bibitem[Dong et al.(2014)]{dong14}
Dong, S.,  Katz, B., Prieto, J. L., Udalski, A., Kozlowski, S. \& Street, R. A.  2014, arXiv1401.1195

\bibitem[Dougherty et al.(1994)]{dougherty94}
Dougherty, S. M., Waters, L. B. F. M., Burki, G., Cote, J., 
Cramer, N., van Kerkwijk, M. H., \& Taylor, A. R. 1994, A\&A, 290, 609

\bibitem[Ekstr{\"o}m et al.(2012)]{Ekstrom12} 
Ekstr{\"o}m, S., Georgy, C.,  Eggenberger, P., et al.\ 2012, A\&A, 537, 146

\bibitem[Graczyk et al.(2003)]{graczyk03}
Graczyk, D.,  Mikolajewski, M., Tomov, T., Kolev, D., \&  Iliev, I.\ 2003, A\&A, 403, 1089

\bibitem[Graczyk et al.(2011)]{graczyk11}
Graczyk, D., Soszy\`nski, I.,  Poleski, R., Pietrzy\`nski, G., Udalski, A. Szyma\`nski, M. K., Kubiak, M.
Wyrzykowski, L. \& Ulaczyk, K. 2011, Acta Astron., 61, 103

\bibitem[Graczyk \& Eyer(2010)]{graczyk10}
Graczyk, D., \& Eyer, L. 2010, Acta Astron., 60, 109	


\bibitem[Galan et al.(2012)]{galan12}
Galan, C., Mikolajewski, M., Tomov, T.,  et al. 2012, A\&A, 544, 53   

\bibitem[Guinan \& DeWarf(2002)]{guinan02}
Guinan, E. F., \& DeWarf, L. E. 2002, in Exotic Stars as Challenges to Evolution,
ed. C. A. Tout, \& W. van Hamme, ASP Conf. Ser., 279, 121

\bibitem[Haisch et al.(2001)]{haisch01}
Haisch, K. E., Jr.,  Lada, E. A., \& Lada, C. J.  2001, ApJ, 553, L153

\bibitem[Halbwachs et al.(2003)]{Halbwachs03}
Halbwachs, J. L., Mayor, M., Udry, S., \& Arenou, F.  2003, A\&A, 397, 159

\bibitem[Harris et al.(2012)]{harris12} 
Harris, R. J., Andrews,  S. M., Wilner, D. J., \& Kraus, A. L.\ 2012, ApJ, 751, 115 

\bibitem[Hedman et al.(2007)]{hedman07}
Hedman, M. M., Nicholson, P. D., Salo, H., Wallis, B. D., Buratti, B. J., 
Baines, K. H., Brown, R. H., \& Clark, R. N. 2007, AJ, 133, 2624

\bibitem[Herbst et al.(2010)]{Herbst10}
Herbst, W., LeDuc, K., Hamilton, C. M., Winn, J. N., Ibrahimov, M., Mundt, R., \& Johns-Krull, C. M.\ 2010, AJ, 140, 2025

\bibitem[Hern\'andez et al.(2007)]{hernandez07}
Hern\'andez, J., Hartmann, L., Megeath, T., Gutermuth, R.,
Muzerolle, J., Calvet, N., Vivas, A. K., Briceno, C., Allen, L., Stauffer, J., Young, E., \& Fazio, G. 2007, ApJ, 662, 1067

\bibitem[Kato et al.(2007)]{kato07}
Kato, D., Nagashima, C., Nagayama, T. et al.\  2007, PASJ, 59, 615
	


\bibitem[Kloppenborg et al.(2010)]{kloppenborg10}
Kloppenborg, B., Stencel, R., Monnier, J. D., et al.\  2010, Nature, 464, 870

\bibitem[Indu \& Subramaniam(2011)]{indu11}
Indu, G., \& Subramaniam, A. 2011, A\&A, 535, 115	 

\bibitem[Mahy et al.(2011)]{mahy11}
Mahy, L., Martins, F., Machado, C., Donati, J.-F., \& Bouret, J.-C. 2011, A\&A, 533, 9

\bibitem[Mamajek et al.(2012)]{mamajek12}
Mamajek, E. E., Quillen, A. C., Pecaut, M. J.,  Moolekamp, F., Scott, E. L., 
Kenworthy, M. A., Collier Cameron, A., \& Parley, N. R. 2012, AJ, 143, 72	

\bibitem[Marigo et al.(2008)]{Marigo08} 
Marigo, P., Girardi, L., Bressan, A., Groenewegen, M. A. T., Silva, L., \& Granato, G. L.\ 2008, A\&A, 482, 883

\bibitem[Massey(2002)]{massey02}
Massey, P.\ 2002, ApJS, 141, 81

\bibitem[Meixner et al.(2006)]{meixner06}
Meixner, M., Gordon, K. D., Indebetouw, R.  et al. 2006, AJ, 132, 2268

\bibitem[Mikolajewski et al.(2005)]{mikolajewski05}
Mikolajewski, M., Galan, C., Gazeas, K., et al.\ 2005, Ap\&SS, 296, 445

\bibitem[Mikolajewski \& Graczyk(1999)]{mikolajewski99}
Mikolajewski, M., \& Graczyk, D. 1999, MNRAS, 303, 521

\bibitem[Pawlak et al.(2013)]{Pawlak13} 
Pawlak, M., Graczyk, D., Soszy{\'n}ski, I., et al.\ 2013, Acta Astron., 63, 323  

\bibitem[Pecaut et al.(2012)]{pecaut12} 
Pecaut, M.~J., Mamajek,  E.~E., \& Bubar, E.~J.\ 2012, \apj, 746, 154 

\bibitem[Pietrukowicz et al.(2013)]{Pietrukowicz13} Pietrukowicz, P., 
Mr{\'o}z, P., Soszy{\'n}ski, I., et al.\ 2013, Acta Astron., 63, 115 

\bibitem[Rivinius et al.(2013)]{rivinius13} 
Rivinius, T., Carciofi, A. C., Martayan, C.	2013, Astron. \& Astrophys. Review, 21, 69	


\bibitem[Scott et al.(2014)]{scott14}
Scott, E. L., Mamajek, E.  E., Pecaut, M., Moolekamp, F., Quillen, A. C., et al.\ 2014, in preparation

\bibitem[Soderblom et al.(2014)]{soderblom14}
Soderblom, D. R., Hillenbrand, L. A., Jeffries, R. D., Mamajek, E. E., \& Naylor, T. 2014,
eprint arXiv:1311.7024,  Ages of young stars,
Accepted for publication as a chapter in Protostars and Planets VI, 
University of Arizona Press (2014), 
eds. H. Beuther, R. Klessen, C. Dullemond, and Th. Henning

\bibitem[Stellingwerf(1978)]{stellingwerf78}
Stellingwerf, R. F. 1978, ApJ, 224, 953




\bibitem[Udalski et al.(2008)]{udalski08}
Udalski, A., Soszynski, I., Szymanski, M., \& Poleski, R.. 2008, Acta Astron., 58, 69

\bibitem[Ulaczyk et al.(2012)]{ulaczyk12}
Ulaczyk, K., Szymanski, M. K., Udalski, A., Kubiak, M., Pietrzynski, G.,
Soszynski, I., Wyrzykowski, L., Poleski, R., Gieren, W., Walker, A., Garcia-Varela, A.\
        Acta Astronomica, 62,  3,  247

\bibitem[Walker(2012)]{Walker12} 
Walker, A.~R.\ 2012, \apss, 341, 43

\bibitem[Waters et al.(1988)]{waters88}
Waters, L.B.F.M., van den Heuvel, E.P.J., Taylor, A.R., Habets, G.M.H.J., \& Persi, P.\ 1988, A\&A, 198, 200 

\bibitem[Wyrzykowski et al.(2003)]{Wyrzykowski03} 
Wyrzykowski, L.,  Udalski, A., Kubiak, M., Szymanski, M., Zebrun, K., Soszynski, I., Wozniak, P. R., 
Pietrzynski, G., \& Szewczyk, O.\ 2003, Acta Astron., 53, 1 

\bibitem[Zaritsky et al.(2004)]{zaritsky04}
Zaritsky, D., Harris, J., Thompson, I. B., Grebel, E. K.\ 2004, AJ, 128, 1606 

\newpage

\end{thebibliography}
\end{document}